\documentclass[twocolumn,showpacs,preprintnumbers,amsmath,amssymb,prl,superscriptaddress]{revtex4-1}
\usepackage{amsfonts}
\usepackage[english]{babel}
\usepackage[T1]{fontenc}
\usepackage{times}
\usepackage{multirow}
\usepackage{mathrsfs}
\usepackage{graphicx}
\usepackage{dcolumn}
\usepackage{bm}
\usepackage[colorlinks,bookmarks=true,citecolor=blue,linkcolor=red,urlcolor=blue]{hyperref}
\usepackage[tight, FIGTOPCAP, hang, raggedright, nooneline]{subfigure}

\newcommand{\bracket}[2]   {  \left<#1 |  #2\right>}
\newcommand{\av}[1]{\langle #1\rangle}

\begin{document}

\title{Fractional Chern Insulators in Topological Flat bands with Higher Chern Number}

\author{Zhao Liu}
\affiliation{Institute of Physics, Chinese Academy of Sciences, Beijing, 100190, China}
\affiliation{Beijing Computational Science Research Center, Beijing, 100084, China}
\author{Emil J. Bergholtz}
\email{Author to whom correspondence should be addressed: ejb@physik.fu-berlin.de}
\affiliation{Dahlem Center for Complex Quantum Systems and Institut f\"ur Theoretische Physik, Freie Universit\"at Berlin, Arnimallee 14, 14195 Berlin, Germany}
\author{Heng Fan}
\affiliation{Institute of Physics, Chinese Academy of Sciences, Beijing, 100190, China}
\author{Andreas M. L\"auchli}
\affiliation{Institut f\"ur Theoretische Physik, Universit\"at Innsbruck, A-6020 Innsbruck, Austria}

\date{\today}

\begin{abstract}
Lattice models forming bands with higher Chern number offer an intriguing possibility for new phases of matter with no analogue in continuum Landau levels. Here, we establish the existence of a number of new bulk insulating states at fractional filling in flat bands with Chern number $C=N>1$, forming in a recently proposed pyrochlore model with strong spin-orbit coupling. In particular, we find compelling evidence for a series of stable states at $\nu=1/(2N+1)$ for fermions as well as bosonic states at $\nu=1/(N+1)$. By examining the topological ground state degeneracies and the excitation structure as well as the entanglement spectrum, we conclude that these states are Abelian. We also explicitly demonstrate that these states are nevertheless qualitatively different from conventional quantum Hall (multilayer) states due to the novel properties of the underlying band structure.
\end{abstract}

\maketitle

{\it Introduction.---} The recent discovery of nearly flat
bands with unit Chern number $C=1$ in itinerant lattice systems \cite{chernins1,chernins2,chernins3} has spurred plenty of theoretical excitement \cite{cherninsnum1,cherninsnum2,bosons,nonab1,nonab2,nonab3,andreas,beyondL,c1a,c1b,c1c,c1d,c1e,qi,bands,cherncf,nonab1,cherntt} as these bands may harbor lattice analogues of fractional quantum Hall (FQH) states that do not require an external magnetic field and may potentially persist at very high temperatures.

While flat $C=1$ bands are very similar to continuum Landau levels, lattice models can harbor bands with higher Chern number and may therefore host qualitatively new phases of matter with no analogue in the continuum. Two independent groups have very recently shown that flat bands with arbitrary higher Chern number can in fact be systematically created in multilayer systems \cite{max,sds} assuming only short-range hopping (see also Ref. \cite{c2} for a nice earlier construction of a flat $C=2$ band) and, as such, provide promising platforms for new phenomena \cite{max,sds,c2,c2num,cn}.

In this work we study the crucial effect of interactions in the original pyrochlore (multi-layer kagome) lattice model proposed for flat bands with arbitrary Chern number $C=N$ \cite{max}. While a very recent study \cite{c2num} focusing on interactions in a fairly flat $C=2$ band on the triangular lattice reported an intriguing $\nu=1/3$ bosonic state, we find compelling evidence for a whole series of incompressible fractional Chern insulator (FCI) states, both for fermions at $\nu=1/(2N+1)$ 
and for bosons at $\nu=1/(N+1)$, $N=2,3,4$.
We also demonstrate that the states which we discover are, despite a number of similarities, qualitatively different from conventional (single and multilayer) FQH states.


{\it Setup.---} We focus on the model describing Rashba spin-orbit coupled particles
on pyrochlore slabs including $N$ kagome layers (from $\mathcal{K}_1$ to $\mathcal{K}_N$) introduced in Ref. \cite{max}.
The single-particle Hamiltonian in real space is
\begin{eqnarray}
H=\sum_{i,j,\sigma}t_{ij}c_{i\sigma}^{\dagger}c_{j\sigma}+\textrm{i}\sum_{i,j,\alpha,\beta}
\lambda_{ij}(\mathbf{E}_{ij}\times \mathbf{R}_{ij})\cdot\mathbf{\sigma}_{\alpha\beta}c_{i\alpha}^{\dagger}c_{j\beta}.
\nonumber
 \label{1pham}
\end{eqnarray}
Although next-nearest hopping is needed to get very flat $C=N$ bands, we consider only nearest neighbor hopping $t_{ij}=t_1=-1$ and $\lambda_{ij}=\lambda_1$ ($\lambda_1=1.1$ for fermions and $\lambda_1=0.9$ for bosons) within each kagome layer, as well as $t_{ij}=t_{\perp}$ \cite{tperp} when involving the triangular layers for simplicity. In momentum space, there are $4N-1$ bands assuming spin polarization, and there is a flat band
with Chern number $C=N$ for suitable choices of hopping parameters \cite{max}.

As is customary we take the flat band limit and consider the case when the flat band is partially filled by interacting particles
with an interaction Hamiltonian $H_\textrm{int}$.
We diagonalize
$H_\textrm{int}$ projected to the flat band for a finite system with $N_1\times N_2$ unit cells.

\begin{figure}
\centerline{\includegraphics[width=\linewidth]{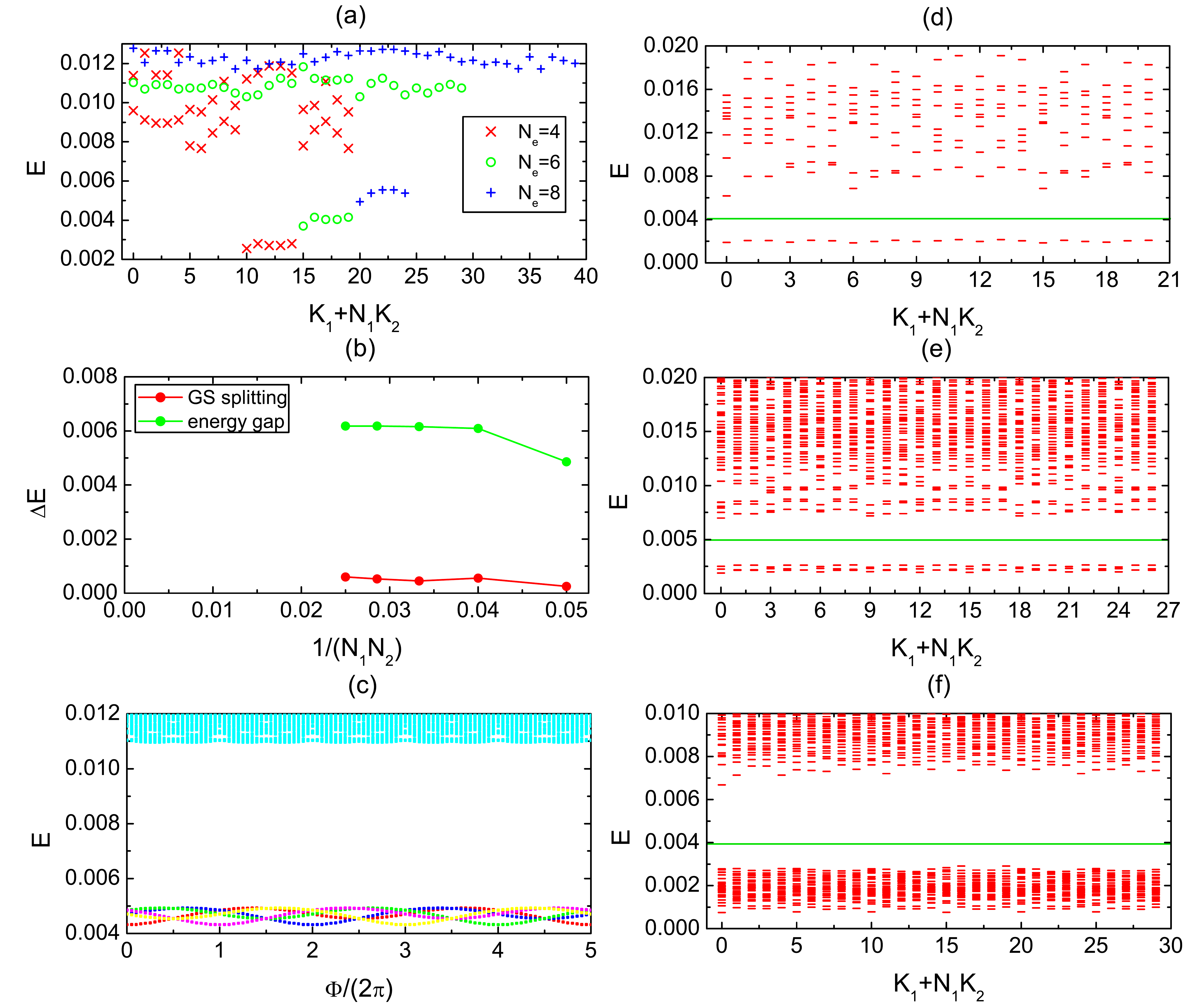}}
\caption{(Color online) Results for the $\nu=1/5, C=2$ fermionic FCI state in a bilayer kagome system with
$N_1=5$ and $N_2=N_e$.
(a) The low-lying energy spectrum for $N_e=4$, $N_e=6$, and $N_e=8$.
(b) The finite-size scaling analysis for both energy gap and ground state splitting.
(c) The $x$-direction spectral flow for $N_e=7$.
(d) The quasihole excitations for $N_e=4$, $N_1=3$, and $N_2=7$ (one hole is added, 21 states below the gap).
(e) The quasihole excitations for $N_e=5$, $N_1=3$, and $N_2=9$ (two holes are added, 81 states below the gap).
(f) The quasihole excitations for $N_e=5$, $N_1=5$, and $N_2=6$ (five holes are added, 756 states below the gap).
\label{fermion_1_5}}
\end{figure}

{\it Fermion energetics.---} We consider fermions with a nearest neighbor repulsion $H_\textrm{int}=\sum_{\av{i,j}}n_in_j$ \cite{layerint} and begin by considering a bilayer kagome system at filling $\nu=1/5$ in the $C=2$ band. For each system size that we study, there are five quasidegenerate ground states at the bottom of the energy spectrum.
These lowest states are separated from other excited states by a clear energy gap, which is a necessary condition
for the $\nu=1/5$ fermionic FCI state [Fig. \ref{fermion_1_5}(a)]. The energy gap is always
significantly larger than the ground state splitting for various system sizes, and a finite-size scaling analysis of the energy gap shows that it is very likely to survive in the thermodynamic limit [Fig. \ref{fermion_1_5}(b)] \cite{splitting}.
To corroborate that the ground states are topologically nontrivial, we also check the spectral flow under twisted boundary conditions, which amounts to inserting magnetic flux through the cycles of the system (i.e., through the handles of the torus). For a many-body state $\Psi$, the twisted boundary condition in the $x(y)$ direction is $\Psi(\mathbf{r}_j+N_{1(2)}\mathbf a_{1(2)})=\exp(i\Phi)\Psi(\mathbf{r}_j)$, where $\Phi$ is the boundary phase and $\mathbf a_{1(2)}$ is the lattice vector. By calculating the spectral flow for some system sizes where the ground states are in different $(K_1,K_2)$ sectors, we find that when $\Phi$ changes from 0 to $5\times 2\pi$, the five ground states evolve into each other, being always separated from excited states by a gap, and finally
return to the initial configuration [Fig. \ref{fermion_1_5}(c)]. The behavior of the spectral flow indicates that the Hall conductance is $\sigma_{H}=\frac{2}{5}\frac{e^2}{h}$ \cite{tknn}, which we have also confirmed by calculating the many-body Chern number.

The quasihole excitations, which carry fractional charge \cite{Laughlin} and statistics \cite{hierarchyHalperin,fstatqh,fstat}, are one of the most important characteristics for FQH states in Landau levels. Here, we first investigate the quasihole excitations of our $C=2$, $\nu=1/5$ fermionic FCI state. By keeping $N_e$ fixed and changing $N_1$ and/or $N_2$, we can add one hole and more holes into the system. A clear gap that separates the low-lying states from high-excited states exists in the quasihole excitation spectrum. The total number of states below the gap is identical to that for the $\nu=1/5$ fermionic Laughlin state in the Landau level on the torus. Moreover, the ground state momenta and counting of quasiholes can be predicted by an exclusion rule known from the thin-torus limit of the FQH system \cite{bk23}: By folding the two-dimensional momenta $(k_1,k_2)$ to one dimension as $k_{{\rm 1D}}\equiv k_1+N_1k_2$ \cite{cherninsnum2}, this rule implies that there are no more than $p$ particles in $q$ consecutive orbitals at, and slightly below, $\nu=p/q$. This shows that the quantum dimension of the quasiholes is $d_{qh}=1$, which is a hallmark of Abelian statistics \cite{eddy}.

\begin{figure}
\centerline{\includegraphics[width=\linewidth]{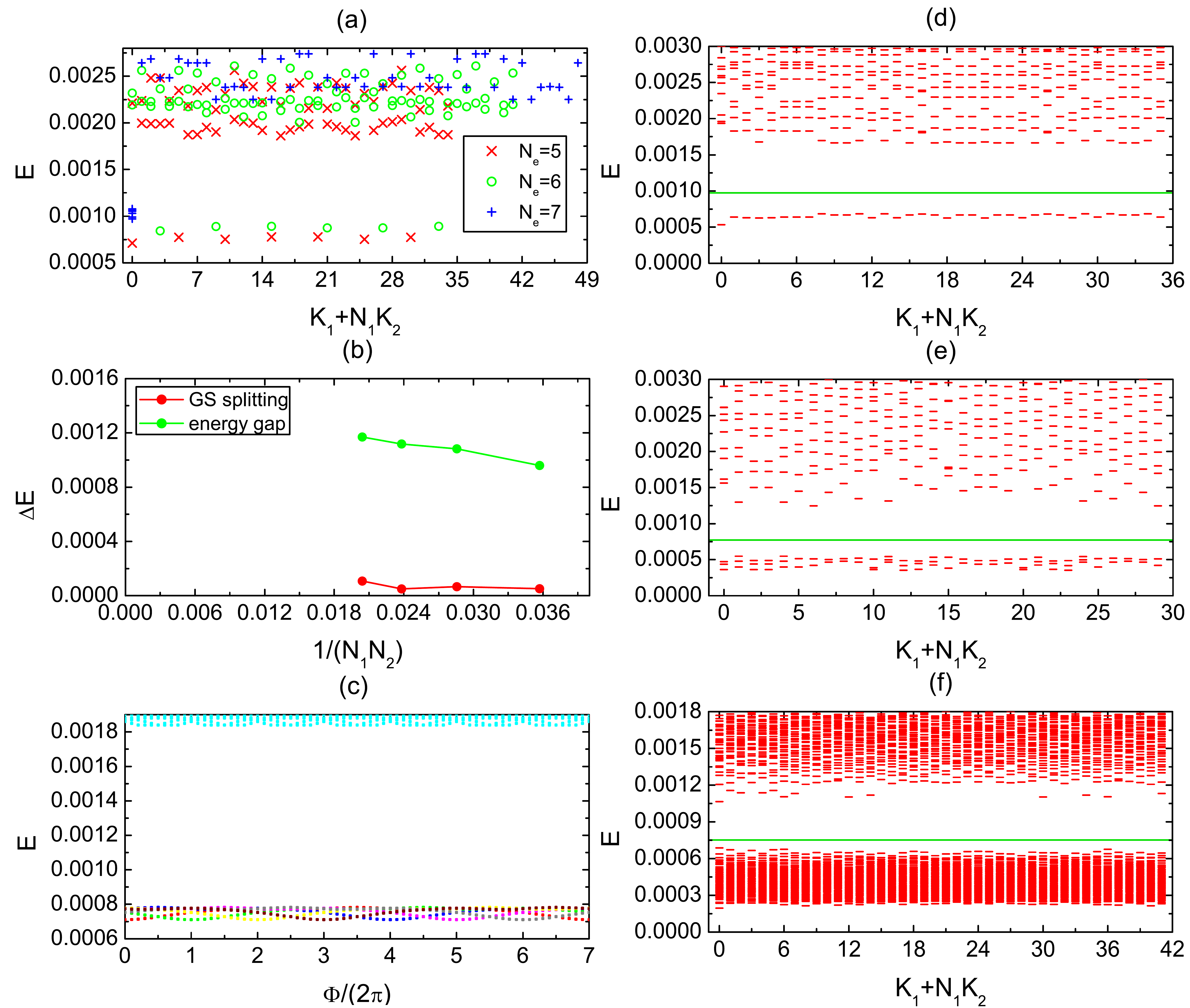}}
\caption{(Color online) Results for the $\nu=1/7, C=3$ fermionic FCI state in a three-layer kagome system with
$N_1=N_e$ and $N_2=7$.
(a) The low-lying energy spectrum for $N_e=5$, $N_e=6$, and $N_e=7$.
(b) The finite-size scaling analysis for both energy gap and ground state splitting.
(c) The $y$-direction spectral flow for $N_e=5$.
(d) The quasihole excitations for $N_e=5$, $N_1=6$, and $N_2=6$ (one hole is added, 36 states below the gap).
(e) The quasihole excitations for $N_e=4$, $N_1=5$, and $N_2=6$ (two holes are added, 75 states below the gap).
(f) The quasihole excitations for $N_e=5$, $N_1=6$, and $N_2=7$ (seven holes are added, 2772 states below the gap).
\label{fermion_1_7}}
\end{figure}

In the three-layer kagome system, we find similar evidence for a $\nu=1/7, C=3$ fermionic FCI state (see
Fig. \ref{fermion_1_7} for a summary of these results).
The gap scaling is again convincing albeit with a gap being roughly $20\%$ of the gap observed at $\nu=1/5$ in the $C=2$ band. A clear gap in the quasihole excitation spectrum of the $\nu=1/7$ state can be identified, and the total number of states below the gap is identical to that for the $\nu=1/7$ fermionic Laughlin state in the Landau level on the torus. In the four-layer kagome system, we find some clues of $\nu=1/9, C=4$ fermionic FCI state, again clearly resolved, albeit with a yet smaller gap \cite{supmat}.

\begin{figure}
\centerline{\includegraphics[width=\linewidth]{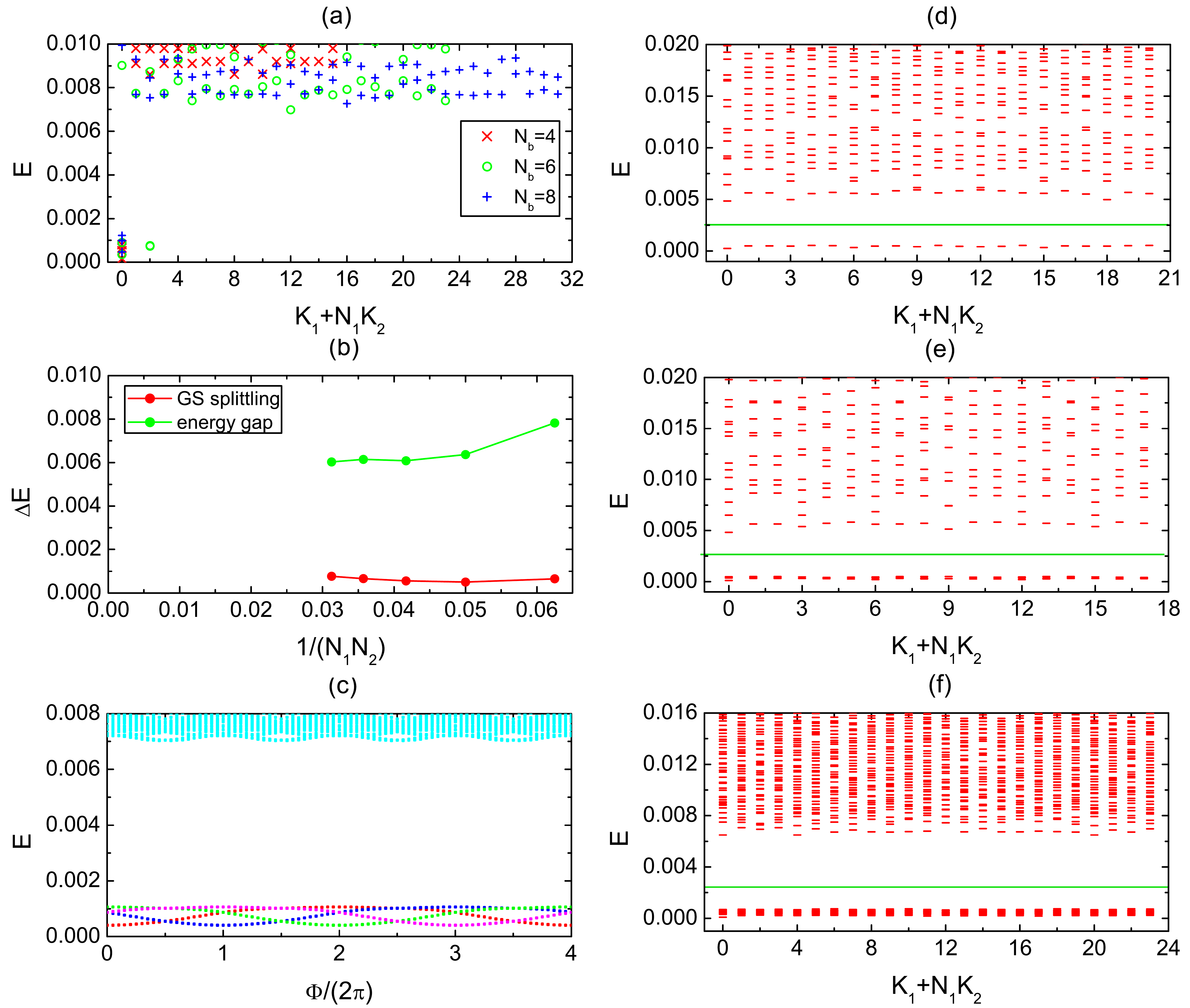}}
\caption{(Color online) Evidence for the $\nu=1/4, C=3$ bosonic FCI state in a three-layer kagome system with
$N_1=4$ and $N_2=N_b$.
(a) The low-lying energy spectrum for $N_b=4$, $N_b=6$, and $N_b=8$.
(b) The finite-size scaling analysis for both energy gap and ground state splitting.
(c) The $x$-direction spectral flow for $N_b=7$.
(d) The quasihole excitations for $N_b=5$, $N_1=3$, and $N_2=7$ (one hole is added, 21 states below the gap).
(e) The quasihole excitations for $N_b=4$, $N_1=3$, and $N_2=6$ (two holes are added, 45 states below the gap).
(f) The quasihole excitations for $N_b=5$, $N_1=4$, and $N_2=6$ (four holes are added, 336 states below the gap).
\label{boson_1_4}}
\end{figure}
{\it Boson energetics.---} We now turn our attention to bosons with on-site repulsion $H_\textrm{int}=\sum_{i}n_i(n_i-1)$
in the bilayer, three-layer, and four-layer kagome systems. In the bilayer kagome system, we find
a $\nu=1/3$ bosonic FCI state  \cite{supmat}, presumably corresponding to the same phase as the $\nu=1/3$ recently observed in a $C=2$ band on the triangular lattice \cite{c2num}.
In the three-layer kagome system, we find convincing evidence for a new $\nu=1/4$ bosonic FCI state in the $C=3$ band as displayed in Fig. \ref{boson_1_4} \cite{quarterfilling}. We also find
an interesting $\nu=1/5, C=4$ bosonic FCI state in the four-layer kagome system \cite{supmat}. Note that, in contrast to $C=1$ systems, bosonic FCI states can form at a Laughlin fraction with an odd denominator when $C>1$. In each case, a clear gap exists above the ground state manifold, as well as in the quasihole excitation spectrum.
Again, the counting of quasiholes can be predicted by the thin-torus-like exclusion rule. While the rule for predicting ground state momenta of the boson states must be slightly modified \cite{hcnonab}, this is analogous with the case in $C=1$ bands. 

\begin{figure}
\centerline{\includegraphics[width=\linewidth]{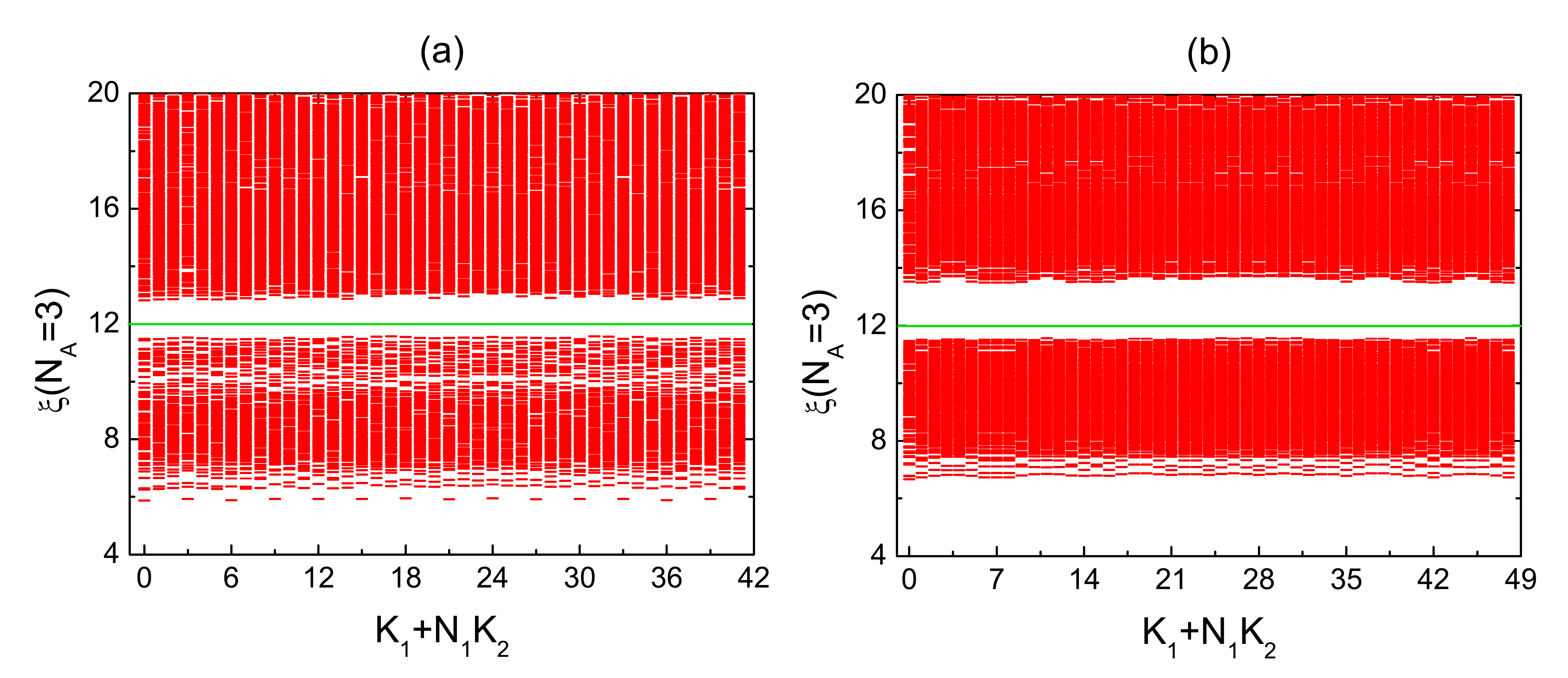}}
\caption{(Color online) Particle entanglement spectra probing the $N_A=3$ quasihole excitations of the (a) $N_e=7, \nu=1/5, C=2$ state on the $5\times7$ lattice (2695 states below the gap) and (b) $N_e=7,\nu=1/7,C=3$ state on the $7\times7$ lattice (7105 states below the gap).
\label{PES}}
\end{figure}

{\it Entanglement spectra.---}
To further corroborate our finding of FCI phases, we have investigated the particle entanglement spectra \cite{LiH,PES}. This provides an independent test of the excitation structure of the system and can be used to discard competing possibilities, such as charge density waves \cite{cherninsnum2}. The results from the largest systems we have studied are displayed in Fig. \ref{PES}. Here, we find that, although the bandwidth of the manifold of low-energy levels is relatively large (which appears to be the generic situation for FCI as well as FQH states), there is a clear entanglement gap separating these levels from generic ones. Strikingly, the number of low-lying levels exactly matches the pertinent quasihole in the corresponding Abelian FQH states. For smaller systems we sometimes observe deviations when $N_A$ is too large \cite{supmat}, but the overall picture conclusively rules out a charge density wave explanation (cf., e.g., Ref. \cite{cherntt}).

\begin{figure}
\centerline{\includegraphics[width=\linewidth]{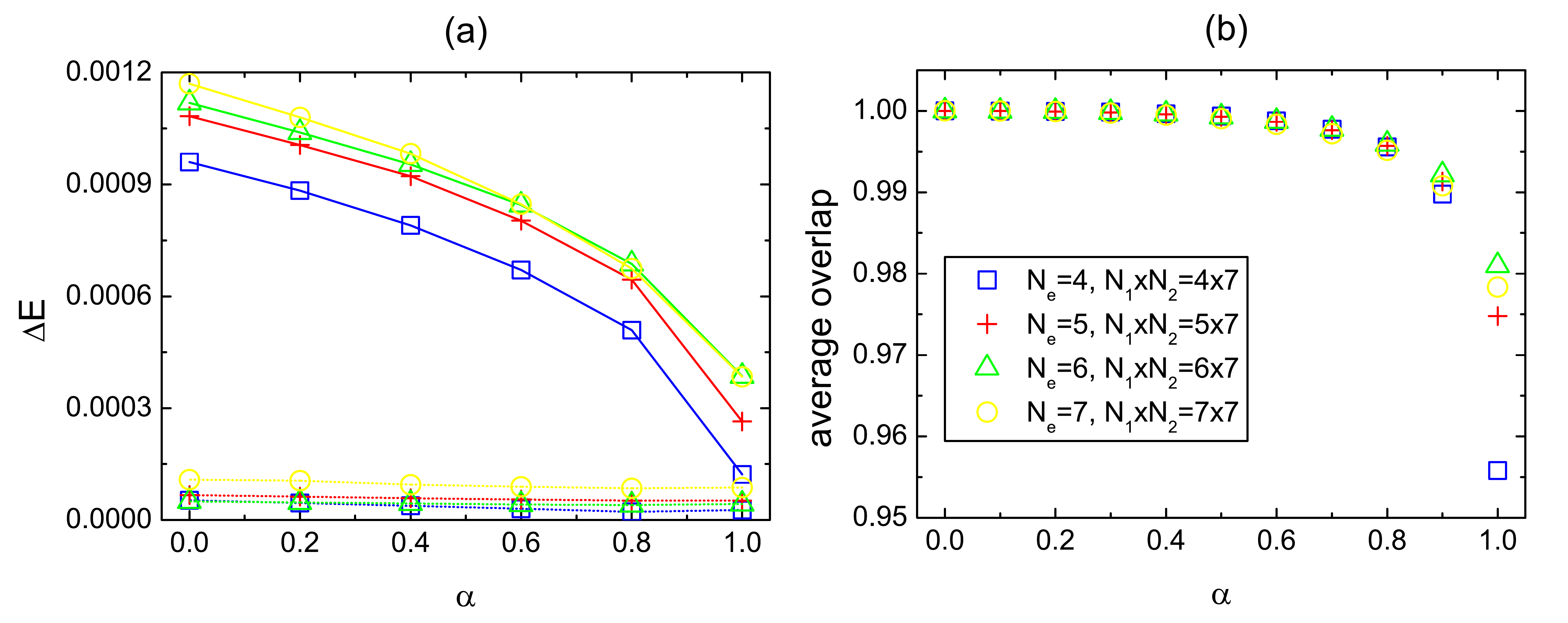}}
\caption{(Color online)
(a) Energy gap (solid line) and ground state splitting (dotted line) as a function of $\alpha$ at $\nu=1/7$
for various system sizes: $N_e=4,N_1=4,N_2=7$ (blue square); $N_e=5,N_1=5,N_2=7$ (red cross);
$N_e=6,N_1=6,N_2=7$ (green triangle); and $N_e=7,N_1=7,N_2=7$ (yellow circle).
(b) The average overlap $\frac{1}{7}\sum_{i=1}^7|\!\bracket{\Psi_i(\alpha)}{\Psi_i(\alpha=0)}\!|$,
where $|\Psi_i(\alpha)\rangle$ is the $i$ th degenerate ground state.
\label{weaken}
}
\end{figure}

{\it Consequences of mutual interlayer entanglement.---}
It is tempting to associate the higher Chern bands with conventional multilayer FQH systems as these systems are topologically equivalent (both have $N$ chiral edge states, etc.). However, the dynamics of all layers in the higher Chern bands is invariably entangled (independent of the interlayer tunneling strength $t_\perp\neq 0$) due to the novel band structure, while the layers in the FQH system can in principle be independent. To probe the consequences of this, we now consider the case of variable interaction strength in the different kagome layers with fermions at $\nu=1/7$ filling in a three-layer kagome system as an example (Fig. \ref{weaken}). Here we weaken the interaction in the midlayer $\mathcal{K}_{2}$ as
$H_\textrm{int}(\alpha)=\sum_{\av{i,j}}n_in_j-\alpha\sum_{\av{i,j}\in \mathcal{K}_{2}}n_in_j$ with $\alpha\in[0,1]$.

We find that, even when the interaction in the midlayer is vanishing ($\alpha=1$),
the ground state degeneracy and energy gap are still finite, and the ground states have a large
overlap with those for $\alpha=0$. We also calculate the spectral flow, the quasihole excitation spectrum, and the entanglement spectrum, and those results confirm that the ground states for $\alpha=1$ are qualitatively the same
as those for $\alpha=0$ \cite{supmat}. This is strikingly different compared to the FQH multilayer case where the gap would close (at least in the weak tunneling regime).

{\it Discussion.---}
In this work, we have provided evidence for a number of new strongly correlated states emerging in nearly flat bands with higher Chern number. These states are likely to be Abelian, as they share ground state degeneracies and excitation structure with Laughlin-like states at the same filling fractions (whenever they exist).
In addition to the fact that the FCI states found here can be realized at anomalous filling fractions (for bosons) compared to single-layer FQH states, we have shown that these new states are also qualitatively distinct from conventional multilayer FQH states by weakening the interaction in the midlayer.

We note that, while multilayer FQH states generically have higher ground state degeneracy than the states found here, it is well known that, e.g., symmetrizing simple Abelian multilayer states can have dramatic consequences \cite{cappelli}, most saliently leading to non-Abelian phases with a reduced ground state degeneracy. Symmetrizing can also lead to Abelian single band FQH states with a yet smaller ground state degeneracy, reminiscent of the FCI states discovered here. In this context, we note that there are a number of intriguing similarities between the states reported here, and composite fermion states for bosons at $\nu=\frac{p}{p+1}$ and fermions at $\nu=\frac{p}{2p+1}$, such as sharing the same Hall conductance, ground state degeneracy, and filling fraction denominator. Moreover, the composite fermion states have the form of (anti)symmetrized multi-layer states, and such states have recently been confirmed to exist also in $C=1$ Chern bands \cite{andreas,beyondL}. This is suggestive of a flux attachment picture \cite{jain89} also for the $C>1$ states presented here.

It is very likely that the $C>1$ bands can also harbor many other incompressible phases, and an exhaustive investigation of the phase diagram remains a challenge for future works using more sophisticated techniques including considering tilted samples \cite{andreas} to obtain a more precise finite-size scaling control. Nevertheless, the states presented here appear to be the most stable ones as indicated by preliminary scans of more generic filling fractions (for relatively small system sizes).

Another open issue is whether our findings are relevant to experiments. One class of candidate systems is cold atom setups with artificial gauge fields \cite{shao,jean,bloch}, while another that is especially well suited for the present model is provided by conveniently grown spin-orbit coupled solid state materials, such as the pyrochlore iridates \cite{pesin,weyl,fiete,kim,pyroqh}.

{\it Note added.---}
Recently, a pre-print with closely related results for bosons, including further $C>1$ FCI states in the model studied here, appeared \cite{hcnonab}.

\acknowledgments

We acknowledge useful correspondence during related collaborations with M. Trescher and R. Moessner. E.J.B. is supported by the Alexander von Humboldt foundation and acknowledges the hospitality of the Institute of Physics at the Chinese Academy of Sciences where parts of this work were carried out.
H. F. is supported by the ``973" program
(Grant No. 2010CB922904).
Z. L. acknowledges MPG RZ Garching for the computational resource.

\newpage
\onecolumngrid

\section*{Supplementary Material for "Fractional Chern Insulators in Topological Flat Bands with Higher Chern Number"}

In this supplementary material for "Fractional Chern Insulators in Topological Flat Bands with Higher Chern Number", we show the energetics of the $\nu=1/9$ fermionic, $\nu=1/7$
fermionic (with weakened interaction in the midlayer), $\nu=1/3$ bosonic and $\nu=1/5$ bosonic FCI states in the multilayer kagome lattice. For several of these systems we also present data on the particle entanglement spectrum (PES). We also give a systematic comparison
between the PES counting of the FCI states and the counting of quasihole excitations derived from a simple exclusion rule (which gives the counting associated with Laughlin states at filling fractions where these exist).

\subsection*{Evidence for the $\nu=1/9, C=4$ fermionic FCI states in the four-layer kagome system}
\begin{figure}[h!]
\centerline{\includegraphics[height=5cm,width=0.7\linewidth]{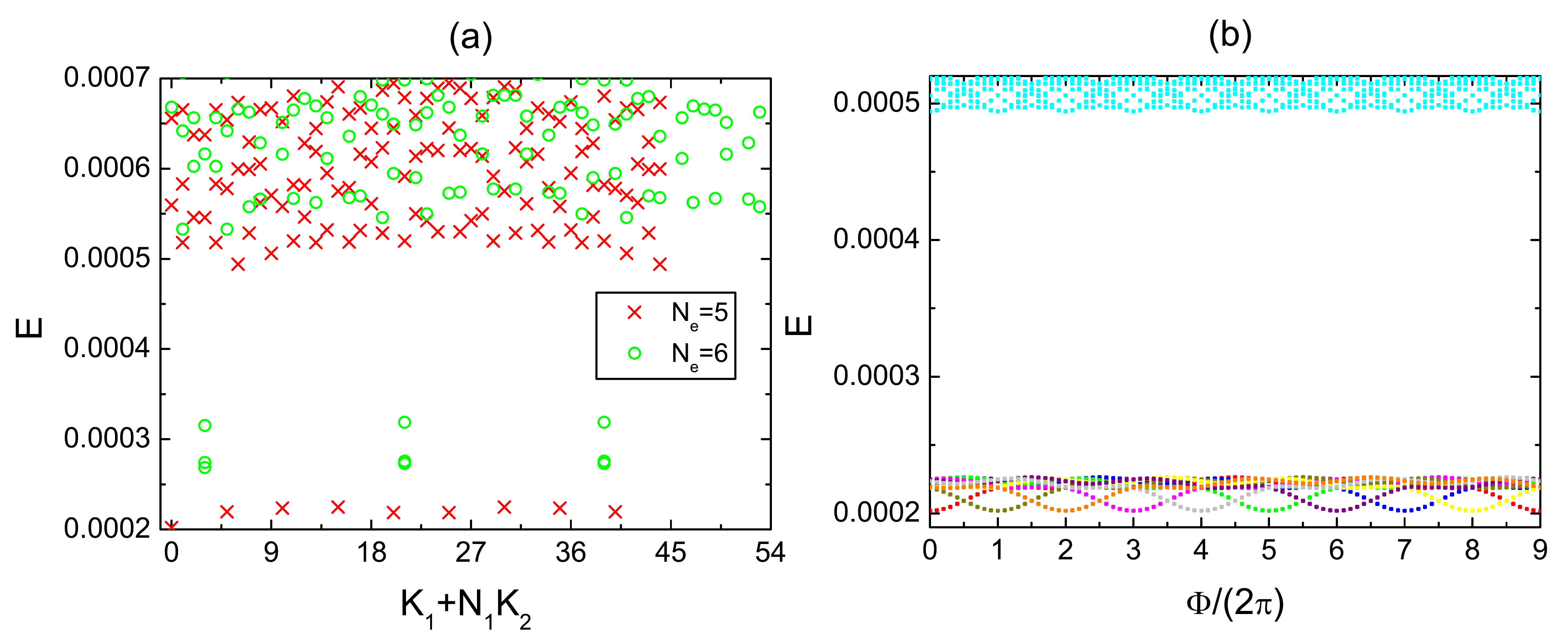}}
\caption{(Color online) Results for the $\nu=1/9$, $C=4$ fermionic FCI states in the four-layer kagome system
($\lambda_1=1.1$). (a) The low-lying energy spectrum for
$N_e=5 (N_1=5,N_2=9)$ and $N_e=6 (N_1=6,N_2=9)$. (b) The $y$-direction spectral flow
for $N_e=5 (N_1=5,N_2=9)$.
\label{fermion_sm}}
\end{figure}

Here we show some data about the existence of the $\nu=1/9$ fermionic FCI states in the four-layer kagome system,
including the energy spectra and spectral flow (Fig. \ref{fermion_sm}).
One can see that there are $9$ nearly degenerate ground states in the energy spectra, which are
separated from other excited states by an energy gap (about $2\times10^{-4}\sim3\times10^{-4}$). The spectral flow under twisted boundary condition
also indicates that the ground states are topologically non-trivial and robust.

\subsection*{Evidence for the $\nu=1/5, C=4$ bosonic FCI states in the four-layer kagome system}
\begin{figure}[h!]
\centerline{\includegraphics[height=12cm,width=0.6\linewidth]{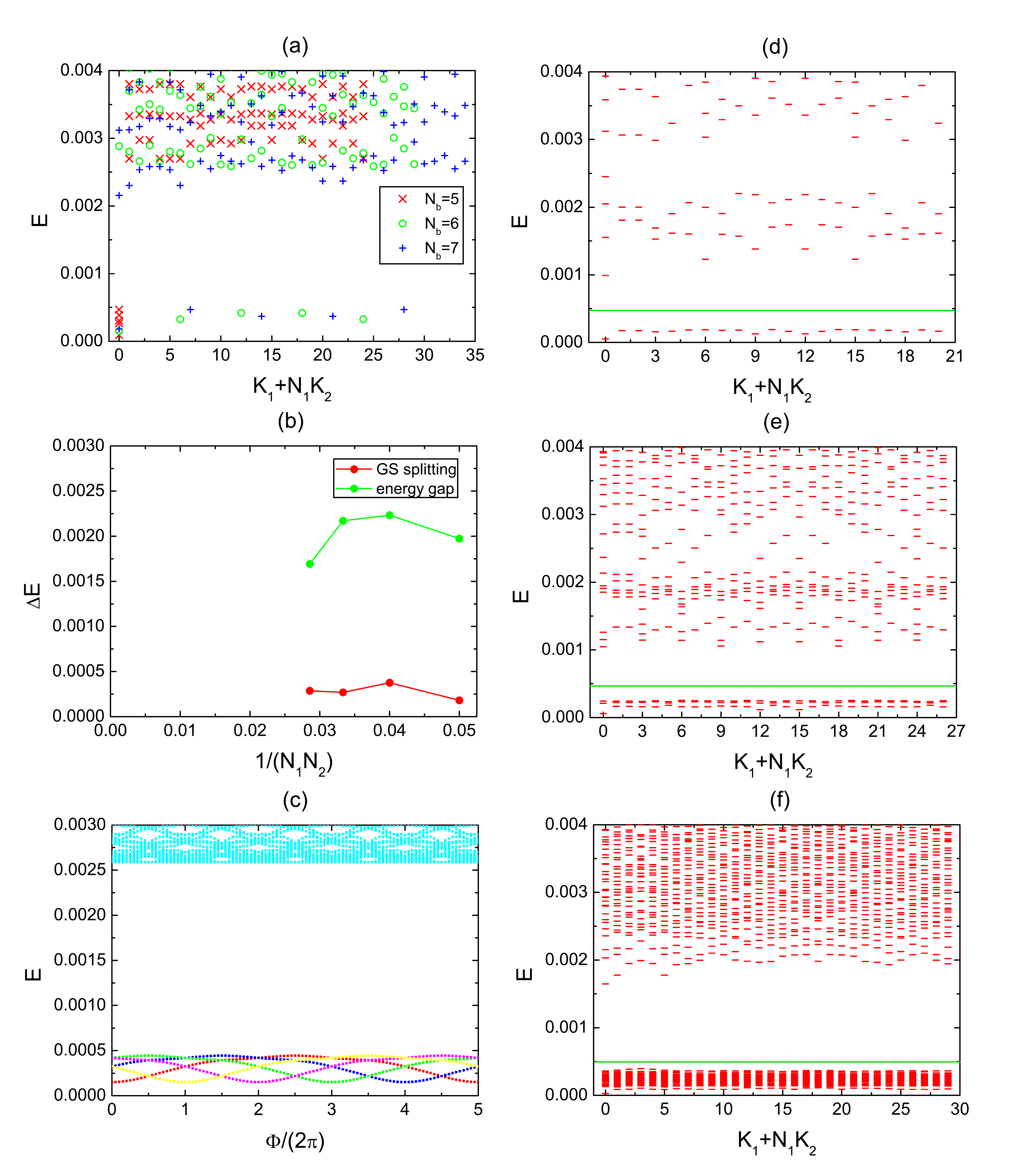}}
\caption{(Color online) Evidence for the $\nu=1/5$, $C=4$ bosonic FCI state in the four-layer kagome system with
$N_2=5$ and $N_1=N_b$ ($\lambda_1=0.9$).
(a) The low-lying energy spectrum for $N_b=5$, $N_b=6$, and $N_b=7$.
(b) The finite-size scaling analysis for both energy gap and ground state splitting.
(c) The $y$-direction spectral flow for $N_b=6$.
(d) $N_b=4$, $N_1=3$, and $N_2=7$ (one hole is added, 21 states below the gap).
(e) $N_b=5$, $N_1=3$, and $N_2=9$ (two holes are added, 81 states below the gap).
(f) $N_b=5$, $N_1=6$, and $N_2=5$ (five holes are added, 756 states below the gap).
\label{boson_1_5}}
\end{figure}

We also find
an interesting $\nu=1/5$ bosonic FCI state in the four-layer kagome system (Fig. \ref{boson_1_5}). This is a new
and somewhat unexpected bosonic state at a Laughlin fraction with odd denominator. A clear gap exists above the ground state manifold, as well as in the quasihole excitation spectrum. The total number of states below the gap is consistent with that for
the $\nu=1/5$ fermionic Laughlin state in Landau level. The energy gap decreases in the last data point in Fig. \ref{boson_1_5}(b). But given the systematics of our findings we still believe that there is an FCI also in this case and that the drop is accidental for this particular system size.

\begin{figure}[h!]
\centerline{\includegraphics[height=5cm,width=0.65\linewidth]{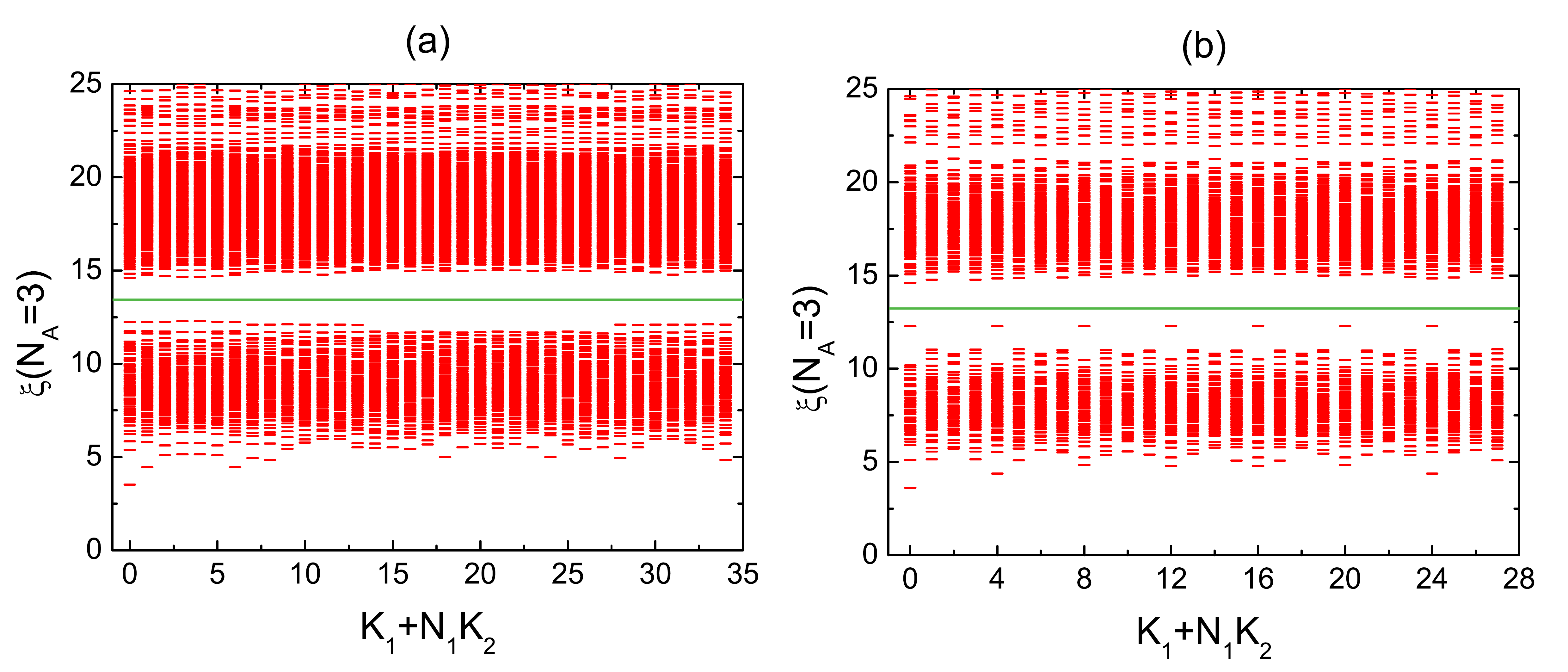}}
\caption{(Color online) (a) The PES for the $\nu=1/5$, $C=4$ bosonic FCI state in the four-layer kagome system with
$N_b=7$, $N_1=7$, and $N_2=5$ (2345 states below the gap).
(b) The PES for the $\nu=1/4$, $C=3$ bosonic FCI state in the three-layer kagome system with
$N_b=7$, $N_1=4$, and $N_2=7$ (1428 states below the gap). $\lambda_1=0.9$.
\label{bosonPES}}
\end{figure}

We also calculate the PES of the $\nu=1/5$ bosonic FCI states (see Fig. \ref{bosonPES}, in which we also
show the PES of the $\nu=1/4$ bosonic FCI states as well for completeness). A clear
entanglement gap can be observed in the PES. A detailed discussion of PES can be found in the last section of this
Supplementary Material.

\subsection*{Evidence for the $\nu=1/3, C=2$ bosonic FCI states in the bilayer kagome system}
\begin{figure}[h!]
\centerline{\includegraphics[height=5cm,width=0.65\linewidth]{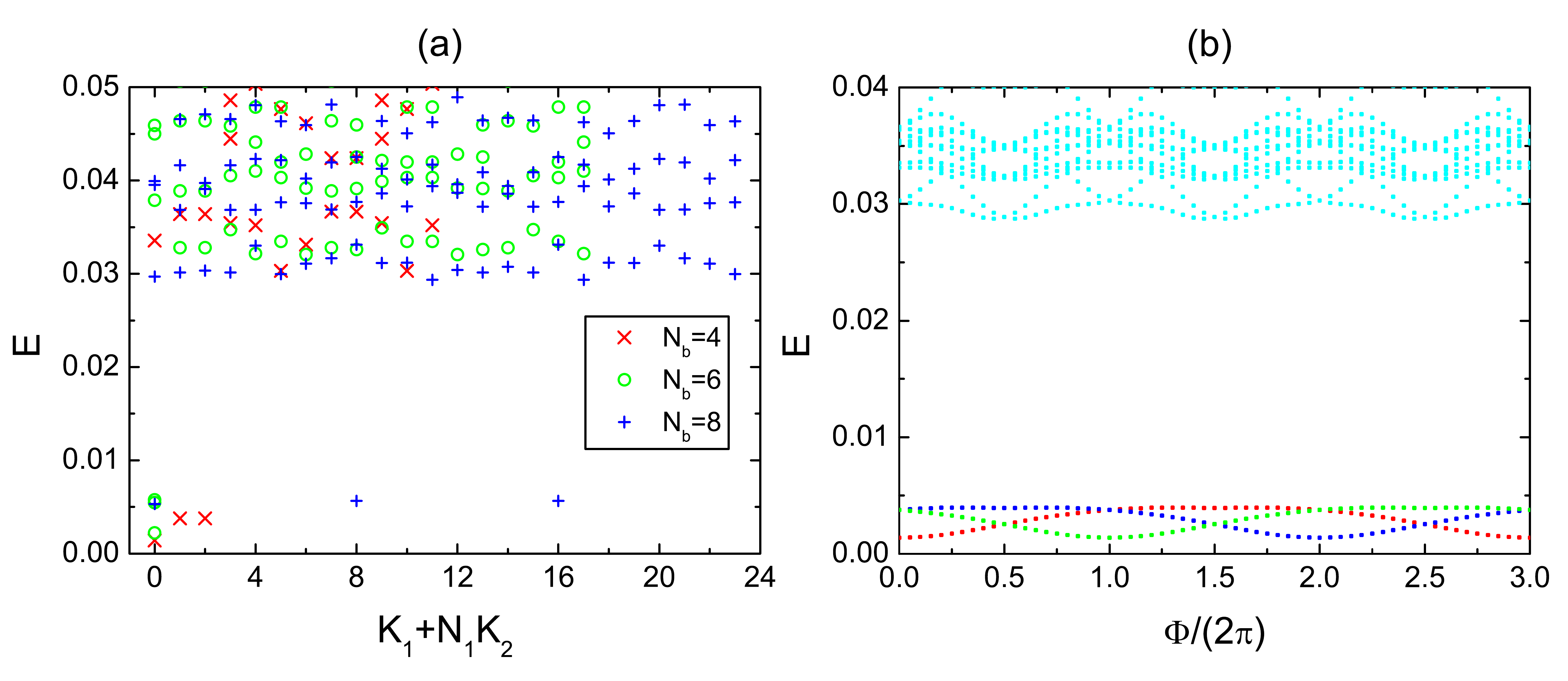}}
\caption{(Color online) Results for the $\nu=1/3$, $C=2$ bosonic FCI states in the bilayer kagome system
($\lambda_1=1$) (a) The low-lying energy spectrum for
$N_b=4 (N_1=3,N_2=4)$, $N_b=6 (N_1=3,N_2=6)$, and $N_b=8 (N_1=4,N_2=6)$. (b) The $x$-direction spectral flow
for $N_b=4 (N_1=3,N_2=4)$.
\label{boson_sm}}
\end{figure}

Bosonic $\nu=1/3$ FCI states were very recently discovered in $C=2$ band in a
triangular lattice model \cite{c2num}. We also observed such states in our
$C=2$ bilayer kagome system model. The ground state degeneracy and spectral flow are shown in Fig. \ref{boson_sm}.

\begin{figure}
\centerline{\includegraphics[width=\linewidth]{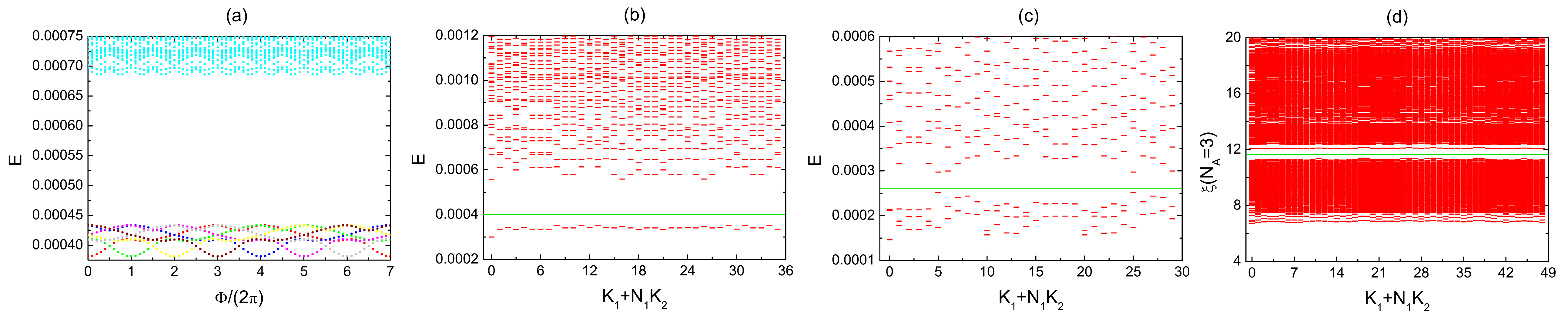}}
\caption{(Color online) Results for the $\nu=1/7$, $C=3$ fermionic FCI states in the three-layer kagome system
($\lambda_1=1.1$) with weakened interaction $H_\textrm{int}(\alpha=1)$, namely
the interaction in the midlayer is eliminated completely. (a) The $y$-direction spectral flow for
$N_e=5 (N_1=5,N_2=7)$. (b) The quasihole excitations for $N_e=5$, $N_1=6$, and $N_2=6$ (one hole is added, 36 states below the gap).
(c) The quasihole excitations for $N_e=4$, $N_1=5$, and $N_2=6$ (two holes are added, 75 states below the gap).
(d) The PES for $N_e=7$, $N_1=7$, and $N_2=7$ (7105 states below the gap).
\label{weaken_sm}}
\end{figure}

\subsection*{Spectral flow, quasihole excitations and entanglement spectrum for $\nu=1/7, C=3$
fermionic FCI states with weakened interaction}
To further confirm that the ground states are FCI states, we calculate the spectral flow, quasihole excitations and PES for $\nu=1/7$
ground states with weakened interaction in the midlayer. We consider an extreme case in which
the interaction in the midlayer is completely eliminated ($\alpha=1$). In Fig. \ref{weaken_sm},
one can find that the spectral flow, quasihole excitation spectra and PES are qualitatively the same
as those for $\alpha=0$.

\subsection*{Data of the PES counting for various filling factors and system sizes}
The particle entanglement spectrum (PES) is usually considered as a valuable tool to
probe the excitation structure of the system and rule out competing states \cite{cherninsnum2}.
For a system possessing a $d$-fold (quasi)degenerate state $\{|\Psi_i\rangle\}_{i=1}^d$, we define the
mixed density matrix of the system as $\rho=\frac{1}{d}\sum_{i=1}^d|\Psi_i\rangle\langle\Psi_i|$.
Then we cut the total particles into two parts $A$ and $B$ with $N_A$ and $N_B$ particles, respectively.
The reduced density matrix of part $A$ can be obtained by tracing out the particles belonging to part $B$,
namely $\rho_A=\textrm{Tr}_B \rho$. We can label each eigenvalue of $\rho_A$ as $e^{-\xi}$, where $\xi$
is just the PES level. As shown in some previous works, the PES of
many FCI states in $C=1$ band has a low-lying part separated from the higher levels by a clear entanglement gap and
the number of levels in the low-lying part (PES counting) matches the quasihole excitations counting of the corresponding FQH states
in Landau level.

In this section, we show the PES counting for both fermionic and bosonic FCI states
that we discover in higher $C$ band and compare it with the quasihole excitations counting provided by the simple
generalized exclusion rule, namely no more than one particle
on $m$ consecutive orbits (see Tab. \ref{t1}):
\begin{eqnarray}
\mathcal{N}_{\textrm{qh}}^m(n,N_{\phi})=\frac{N_{\phi}[N_{\phi}-(m-1)n-1]!}{n!(N_{\phi}-mn)!},
\label{FQHcounting}
\end{eqnarray}
where $N_{\phi}$ is the number of flux (orbits) and $n$ is the number of particles. (This counting also holds for the $\nu=1/m$ Laughlin state on the torus.)

After extracting PES from the FCI state at $\nu=1/m$, we can get its counting $\mathcal{N}_{\textrm{PES}}^m(N_A, N_1, N_2)$.
Then, we want to compare $\mathcal{N}_{\textrm{PES}}^m(N_A, N_1, N_2)$ with
$\mathcal{N}_{\textrm{qh}}^m(N_A,N_{\phi}=N_1N_2)$. In $C=1$ band, they are found to always match \cite{cherninsnum2}.
Although the over all correspondence is striking in our higher $C$ band simulations, we find some deviations. With the present data we cannot conclusively decide if these deviations have a physical meaning or if they are merely due to finite size effects. For fermionic $\nu=1/5$ FCI states in $C=2$ band, $\mathcal{N}_{\textrm{PES}}^m(N_A, N_1, N_2)=\mathcal{N}_{\textrm{qh}}^m(N_A,N_{\phi}=N_1N_2)$. However, at other filling factors, in some cases where the system size is
small and (or) $N_A$ is large, $\mathcal{N}_{\textrm{PES}}^m(N_A, N_1, N_2)$ can be smaller than $\mathcal{N}_{\textrm{qh}}^m(N_A,N_{\phi}=N_1N_2)$.
Despite of this derivation, our results clearly point to the Abelian nature of those FCI states and
rule out the possibility of charge density wave states (for which the PES counting is expected to be significantly smaller \cite{cherntt}).

\begin{table}
\caption{\label{t1} In this table, we give the PES counting of the FCI states
for various filling factors and system sizes. For comparison, we also show the counting of the quasihole
excitations provided by the simple
generalized exclusion rule.}
\centering
\subtable{
\begin{tabular}{|c|c|c|}
\hline
\multicolumn{3}{|c|}{fermions at $\nu=1/5$ in $C=2$ band}\\
\hline
system size ($N_e$, $N_1\times N_2$)&$\mathcal{N}_{\textrm{PES}}^m(N_A, N_1, N_2)$&$\mathcal{N}_{\textrm{qh}}^m(N_A,N_{\phi}=N_1N_2)$\\
\hline
$N_e=4$, $N_1\times N_2=5\times4$&$N_A=2:110$&$N_A=2:110$\\
\hline
$N_e=5$, $N_1\times N_2=5\times5$&$N_A=2:200$&$N_A=2:200$\\
\hline
\multirow{2}{*}{$N_e=6$, $N_1\times N_2=5\times6$}&
$N_A=2:315$& $N_A=2:315$ \\
\cline{2-3}
&$N_A=3:1360$& $N_A=3:1360$ \\
\hline
\multirow{2}{*}{$N_e=7$, $N_1\times N_2=5\times7$}&
$N_A=2:455$& $N_A=2:455$ \\
\cline{2-3}
&$N_A=3:2695$& $N_A=3:2695$ \\
\hline
\multirow{2}{*}{$N_e=8$, $N_1\times N_2=5\times8$}&
$N_A=2:620$& $N_A=2:620$ \\
\cline{2-3}
&$N_A=3:4680$& $N_A=3:4680$ \\
\cline{2-3}
&$N_A=4:17710$& $N_A=4:17710$ \\
\hline
\end{tabular}
}
\subtable{
\begin{tabular}{|c|c|c|}
\hline
\multicolumn{3}{|c|}{fermions at $\nu=1/7$ in $C=3$ band}\\
\hline
system size ($N_e$, $N_1\times N_2$)&$\mathcal{N}_{\textrm{PES}}^m(N_A, N_1, N_2)$&$\mathcal{N}_{\textrm{qh}}^m(N_A,N_{\phi}=N_1N_2)$\\
\hline
$N_e=4$, $N_1\times N_2=4\times7$&$N_A=2:210$&$N_A=2:210$\\
\hline
$N_e=5$, $N_1\times N_2=5\times7$&$N_A=2:385$&$N_A=2:385$\\
\hline
\multirow{2}{*}{$N_e=6$, $N_1\times N_2=6\times7$}&
$N_A=2:609$& $N_A=2:609$ \\
\cline{2-3}
&$N_A=3:3248$& $N_A=3:3542$ \\
\hline
\multirow{2}{*}{$N_e=7$, $N_1\times N_2=7\times7$}&
$N_A=2:882$& $N_A=2:882$ \\
\cline{2-3}
&$N_A=3:7105$& $N_A=3:7105$ \\
\hline
\end{tabular}
}
\subtable{
\begin{tabular}{|c|c|c|}
\hline
\multicolumn{3}{|c|}{bosons at $\nu=1/4$ in $C=3$ band}\\
\hline
system size ($N_b$, $N_1\times N_2$)&$\mathcal{N}_{\textrm{PES}}^m(N_A, N_1, N_2)$&$\mathcal{N}_{\textrm{qh}}^m(N_A,N_{\phi}=N_1N_2)$\\
\hline
$N_b=4$, $N_1\times N_2=4\times4$&$N_A=2:72$&$N_A=2:72$\\
\hline
$N_b=5$, $N_1\times N_2=4\times5$&$N_A=2:130$&$N_A=2:130$\\
\hline
\multirow{2}{*}{$N_b=6$, $N_1\times N_2=4\times6$}&
$N_A=2:204$& $N_A=2:204$ \\
\cline{2-3}
&$N_A=3:680$& $N_A=3:728$ \\
\hline
\multirow{2}{*}{$N_b=7$, $N_1\times N_2=4\times7$}&
$N_A=2:294$& $N_A=2:294$ \\
\cline{2-3}
&$N_A=3:1428$& $N_A=3:1428$ \\
\hline
\multirow{2}{*}{$N_b=8$, $N_1\times N_2=4\times8$}&
$N_A=2:400$& $N_A=2:400$ \\
\cline{2-3}
&$N_A=3:2464$& $N_A=3:2464$ \\
\cline{2-3}
&$N_A=4:7088$& $N_A=4:7752$ \\
\hline
\end{tabular}
}
\subtable{
\begin{tabular}{|c|c|c|}
\hline
\multicolumn{3}{|c|}{bosons at $\nu=1/5$ in $C=4$ band}\\
\hline
system size ($N_b$, $N_1\times N_2$)&$\mathcal{N}_{\textrm{PES}}^m(N_A, N_1, N_2)$&$\mathcal{N}_{\textrm{qh}}^m(N_A,N_{\phi}=N_1N_2)$\\
\hline
$N_b=4$, $N_1\times N_2=4\times5$&$N_A=2:90$&$N_A=2:110$\\
\hline
$N_b=5$, $N_1\times N_2=5\times5$&$N_A=2:200$&$N_A=2:200$\\
\hline
\multirow{2}{*}{$N_b=6$, $N_1\times N_2=6\times5$}&
$N_A=2:315$& $N_A=2:315$ \\
\cline{2-3}
&$N_A=3:1000$& $N_A=3:1360$ \\
\hline
\multirow{2}{*}{$N_b=7$, $N_1\times N_2=7\times5$}&
$N_A=2:455$& $N_A=2:455$ \\
\cline{2-3}
&$N_A=3:2345$& $N_A=3:2695$ \\
\hline
\end{tabular}
}
\end{table}


\begin{thebibliography}{99}


\bibitem{chernins1}E. Tang, J.-W. Mei, and X.-G. Wen,
Phys. Rev. Lett. {\bf 106}, 236802 (2011).

\bibitem{chernins2}K. Sun, Z. Gu, H. Katsura, and S. Das Sarma,
Phys. Rev. Lett. {\bf 106}, 236803 (2011).

\bibitem{chernins3}T. Neupert, L. Santos, C. Chamon, and C. Mudry,
Phys. Rev. Lett. {\bf 106}, 236804 (2011).

\bibitem{cherninsnum1}D. N. Sheng, Z. Gu, K. Sun, and L. Sheng, Nat. Commun. {\bf 2}, 389 (2011).

\bibitem{cherninsnum2}N. Regnault and B. A. Bernevig,  Phys. Rev. X {\bf 1}, 021014 (2011).

\bibitem{bosons}Y.-F. Wang, Z.-C. Gu, C.-D. Gong, and D. N. Sheng,  Phys. Rev. Lett. {\bf 107}, 146803 (2011).

\bibitem{nonab1}B. A. Bernevig and N. Regnault, Phys. Rev. B {\bf 85}, 075128 (2012).

\bibitem{nonab2} Y.-F. Wang, H. Yao, Z.-C. Gu, C.-D. Gong, and D. N. Sheng,  Phys. Rev. Lett. {\bf 108}, 126805 (2012).

\bibitem{nonab3} Y.-L. Wu, B. A. Bernevig, and N. Regnault, Phys. Rev. B {\bf 85}, 075116 (2012).

\bibitem{beyondL} T. Liu, C. Repellin, B. A. Bernevig, and N. Regnault, arXiv:1206.2626.

 \bibitem{andreas} A. L\"auchli, Z. Liu, E.J. Bergholtz, and R. Moessner, arXiv:1207.6094.

 \bibitem{c1a}J.W.F. Venderbos, M. Daghofer, and J. van den Brink, Phys. Rev. Lett. {\bf 107}, 116401 (2011).

 \bibitem{c1b} X. Hu, M. Kargarian, and G. A. Fiete,  Phys. Rev. B {\bf 84}, 155116 (2011).

 \bibitem{c1c} D. Xiao, W. Zhu, Y. Ran, N. Nagaosa, and S. Okamoto, Nat. Commun. {\bf 2},  596 (2011).

\bibitem{c1d} J.W.F. Venderbos, S. Kourtis, J. van den Brink, and M. Daghofer, Phys. Rev. Lett. {\bf 108}, 126405 (2012).

\bibitem{c1e} P. Ghaemi, J. Cayssol, D. N. Sheng, and A. Vishwanath, Phys. Rev. Lett. {\bf 108}, 266801 (2012).

 \bibitem{qi}X.-L. Qi, Phys. Rev. Lett. {\bf 107}, 126803 (2011).

\bibitem{bands}S.~A.~Parameswaran, R.~Roy, and S.~L.~Sondhi, Phys. Rev. B {\bf 85}, 241308(R) (2012); M.~O.~Goerbig, Eur.~Phys.~J.~B {\bf 85}, 15 (2012).

\bibitem{cherncf} A. Vaezi, arXiv:1105.0406;  G. Murthy and R. Shankar, arXiv:1108.5501; J. McGreevy, B. Swingle, and K.-A. Tran, Phys. Rev. B {\bf 85}, 125105 (2012); Y.-M. Lu and Y. Ran, Phys. Rev. B {\bf 85}, 165134 (2012).

 \bibitem{cherntt} B. A. Bernevig and N. Regnault, arXiv:1204.5682.


\bibitem{max}M. Trescher and E.J. Bergholtz, arXiv:1205.2245.

\bibitem{sds}
S. Yang, Z.-C. Gu, K. Sun, and S. Das Sarma, arXiv:1205.5792.

\bibitem{c2} F. Wang and Y. Ran, Phys. Rev. B {\bf 84}, 241103(R) (2011).


\bibitem{c2num} Y.-F. Wang, H. Yao, C.-D. Gong, and D. N. Sheng, arXiv:1204.1697.


\bibitem{cn}M. Barkeshli and X.-L. Qi, Phys. Rev. X {\bf 2}, 031013 (2012).


\bibitem{tperp} The properties of the $C=N$ band, and hence the projected band interaction, do not depend on $t_{\perp}$ (as long as $t_{\perp}\neq0$ which is needed for the formation of the $C=N$ band).


\bibitem{layerint}
Identical results are obtained if the interaction is restricted to the kagome layers, as the wave functions have a vanishing amplitude on the triangular lattice sites.

\bibitem{splitting}
In this initial study we keep $N_1=5$ fixed and change $N_2$ (or keep $N_2=5$ fixed and change $N_1$), so we cannot see an exponential decay of the ground state splitting.

	
\bibitem{tknn}D. J. Thouless, M. Kohmoto, M. P. Nightingale, and M. den Nijs, Phys. Rev. Lett. {\bf 49}, 405 (1982).

 \bibitem{Laughlin}
	 R.~B.~Laughlin,
	 Phys. Rev. Lett. {\bf 50}, 1395 (1983).

\bibitem{hierarchyHalperin} B.I. Halperin, Phys. Rev. Lett. {\bf 52}, 1583, 2390(E)
(1984).

\bibitem{fstatqh}D. Arovas, J. R. Schrieffer, and F. Wilczek, Phys. Rev. Lett. {\bf 53}, 722 (1984).

\bibitem{fstat}J. M. Leinaas and J. Myrheim, Nuovo Cimento Soc. Ital. Fis. {\bf 37B}, 1 (1977).

\bibitem{bk23}
	 E.~J.~Bergholtz and A.~Karlhede, J.~Stat.\ Mech.\ (2006) L04001;
	 Phys.\ Rev.\ B {\bf 77}, 155308 (2008).

\bibitem{eddy} For a pedagogical account of the relation between exclusion rules and FQH states, see E. Ardonne, E. J. Bergholtz, J. Kailasvuori, and E. Wikberg, J. Stat. Mech. (2008) P04016.

    \bibitem{supmat} See
the Supplementary Material for further details.

\bibitem{quarterfilling}We note that a possible $\nu=1/4$ bosonic FCI state in another $C=2$ band was discussed in Ref. \cite{c2num}. However, in that case, the degeneracy of ground states strongly depends on the system size,
and the many-body Chern number showed anomalous behavior. We do not find a similar state in our $C=2$ band, and the $\nu=1/4$ bosonic FCI state in our $C=3$ band is also clearly different from the unexplained state of Ref. \cite{c2num}.

\bibitem{hcnonab} A. Sterdyniak, C. Repellin, B. Andrei Bernevig, and N. Regnault, arXiv:1207.6385.
	
\bibitem{LiH}H. Li and F. D. M. Haldane, Phys. Rev. Lett. 101, 010504
(2008).
\bibitem{PES} A. Sterdyniak, N. Regnault, and B. A. Bernevig, Phys.
Rev. Lett. 106, 100405 (2011).

\bibitem{cappelli}
A. Cappelli, L.S. Georgiev, and I.T. Todorov,
Nucl. Phys. B {\bf 599}, 499 (2001).

\bibitem{jain89} J.K. Jain,  Phys. Rev. Lett. {\bf 63}, 199 (1989).

\bibitem{shao}
L. B. Shao, S.-L. Zhu, L. Sheng, D. Y. Xing, and Z. D. Wang, Phys. Rev. Lett.
{\bf 101}, 246810 (2008).

\bibitem{jean}
J. Dalibard, F. Gerbier, G. Juzeliunas, and P. Ohberg, Rev. Mod. Phys. {\bf 83}, 1523 (2011).

\bibitem{bloch}
M. Aidelsburger, M. Atala, S. Nascimbene, S. Trotzky, Y.-A. Chen, and I. Bloch,
Phys. Rev. Lett. {\bf 107}, 255301 (2011).

\bibitem{pesin} D. A. Pesin and L. Balents, Nature Phys. {\bf 6}, 376 (2010).

\bibitem{weyl}X. Wan, A. M. Turner, A. Vishwanath, and S. Y. Savrasov, Phys. Rev. B {\bf 83}, 205101 (2011).

\bibitem{fiete}M. Kargarian, J. Wen, and G. A. Fiete, Phys. Rev. B {\bf 83}, 165112 (2011).

\bibitem{kim} W. Witczak-Krempa and Y. B. Kim, Phys. Rev. B {\bf 85}, 045124 (2012).

\bibitem{pyroqh}Y. Machida, S. Nakatsuji, Y. Maeno, T. Tayama, T. Sakakibara, and
S. Onoda, Phys. Rev. Lett. {\bf 98}, 057203 (2007).


\end{thebibliography}
\end{document}